\documentclass{desyproc}

\begin{document}
\title{Heavy quark production at the LHC \\in the Parton Reggeization Approach}

\author{{\slshape Anton Karpishkov, Maxim Nefedov, \underline{Vladimir Saleev}, Alexandra Shipilova}\\[1ex]
Samara National Research University, Moskovskoe Shosse, 34, 443086,
Samara, Russia}

\contribID{xy}

\confID{999}  
\desyproc{DESY-PROC-2099-01}
\acronym{HQSF-2016} 
\doi  

\maketitle

\begin{abstract}
We present the general introduction to the Parton Reggeization
Approach and discuss some new results on large-$p_T$ production for
heavy quarks at the LHC.  We concentrate here on study of different
correlation spectra for $b\bar b-$jet pairs,  $D\bar D-$meson pairs
and $DD-$meson pairs.
\end{abstract}

\section{Introduction}

  Study of the production of mesons, containing $c$ and $b$ quarks in hadronic collisions, persists as a topic of
  considerable interest both in theoretical and experimental communities of high energy physics. Production of
  the heavy-quark jets and $D$ or $B$ mesons is a good testing ground for the advanced computational
techniques in perturbative Quantum Chromodynamics (pQCD),
  because the mass of the heavy quark provides a lower bound for the hard scale of the process, $\mu \geqslant 2 m_{c/b} \gg \Lambda_{QCD}$.
  Therefore, one can expect that higher-order perturbative corrections in $\alpha_s(\mu)$ can be taken under control. On the other hand,
  due to the large center-of-mass energy of the modern hadron colliders, most notably the LHC, $c$ and $b$ quarks are produced copiously.
  Huge statistics allows experimental collaborations to perform the measurements of a large variety of {\it correlation} observables.
  These observables are differential in such variables as invariant mass of a pair of reconstructed hadrons ($M$), azimuthal angle between
  their transverse momenta ($\Delta\phi$) or distance in rapidity between the reconstructed hadrons ($\Delta y$).

  Due to the nontrivial kinematical cuts on the phase-space of reconstructed hadrons, these correlation observables become multi-scale quantities.
  The conventional formalism of the Collinear Parton Model (CPM) is most suitable for the calculation of single-scale observables,
  and therefore leads to a significant radiative corrections for the above mentioned correlation spectra.
  The aim of Parton Reggeization Approach (PRA), which will be described below, is to introduce the gauge-invariant scheme of QCD-factorization,
  which will take into account the leading part of higher-order corrections, which are needed to describe the above mentioned correlation observables,
  already in the leading order (LO), and to improve, in such a way, order-by-order stability of the predictions.

  Moreover, in many measurements, the heavy hadrons are produced with significant transverse momentum $p_T\gg m_{c/b}$.
  In this kinematics, the higher-order corrections are enhanced by the "fragmentation" logarithms  $\log (p_T/m_{c/b})$,
  which should be resummed into the scale-dependent parton to hadron Fragmentation Functions (FF) $F_{i \to {\cal H}}(z, \mu^2)$,
  where $z$ is the fraction of the momentum of the parent parton $i$, which is carried by the hadron ${\cal H}$.
  Not only the $c/b\to {\cal H}_{c/b}$ fragmentation function become enhanced, but also a sizable $g\to {\cal H}_{c/b}$ fragmentation function
   is generated by the Dokshitzer-Gribov-Lipatov-Altarelli-Parisi (DGLAP) evolution of fragmentation functions. The gluon channel has to be taken into account due to the huge cross-section
   of the production of gluons
    at hadron colliders.


  The outline of the current contribution is the following: in the Sec.~\ref{sec:LO-PRA} the general scheme of LO PRA calculations is described,
  in the Sec.~\ref{sec:heavy} some new results on the heavy-flavor production, obtained in the LO of PRA are outlined.

\section{Parton Reggeization Approach at the Leading Order \label{sec:LO-PRA}}
  The standard alternative to the exact order-by-order treatment of the additional radiation within the CPM is the method of QCD Monte-Carlo (MC)
   generators, such as PYTHIA or HERWIG. The latter is based on the MC-simulation of DGLAP evolution and the property of collinear factorization
    of matrix elements in QCD, which allows one to take into account the radiation of additional partons with $p_T\ll \mu$ within the
    Leading Logarithmic Approximation (LLA). Strictly speaking, this method is inapplicable for the simulation of radiation of additional
    partons with $p_T\sim \mu$, since collinear factorization of QCD matrix elements is not valid in this region. Above-mentioned correlation
    observables are strongly sensitive to the radiation of additional hard partons, and description of this observables therefore depends
    on the {\it ad-hoc} scheme of treatment of the hard radiation,
    \footnote{Such scheme usually includes boosts and rotations of the matrix element with {\it on-shell} initial-state partons
    and the so-called recoiling scheme. The latter restores overall momentum conservation in the event.}
    which is implemented in particular MC generator.

  To improve the collinear approximation in the region of $p_T\sim \mu$, PRA relies on another factorization theorem for QCD matrix elements,
  namely, factorization in the Multi-Regge Kinematics (MRK), see Ref.~ \cite{LipatovRev}  for the review. In the MRK, the final state partons
  may be grouped into clusters, which are highly separated in rapidity, but the transverse momenta of partons in this clusters may be arbitrarily high.
  LO calculations in the PRA are based on the certain approximate expression for QCD matrix element with the emission of two partons
  additionally to ones, required by the LO hard subprocess. This approximate expression smoothly interpolates between collinear and Regge
  limits for the emission of additional partons, and allows us to improve the description of emissions of additional hard partons on the solid
  theoretical grounds. Below, we will briefly outline the main steps of derivation of the factorization formula of the PRA for
  gluon-initiated processes.

\begin{figure}[h]
\begin{center}
\includegraphics[width=0.3\textwidth]{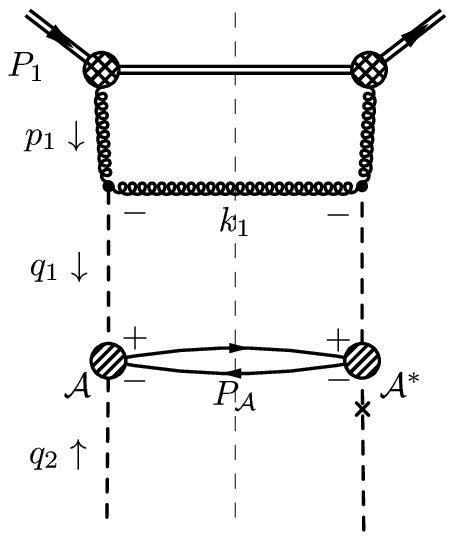}\includegraphics[width=0.4\textwidth, angle=-90,origin=c, clip=]{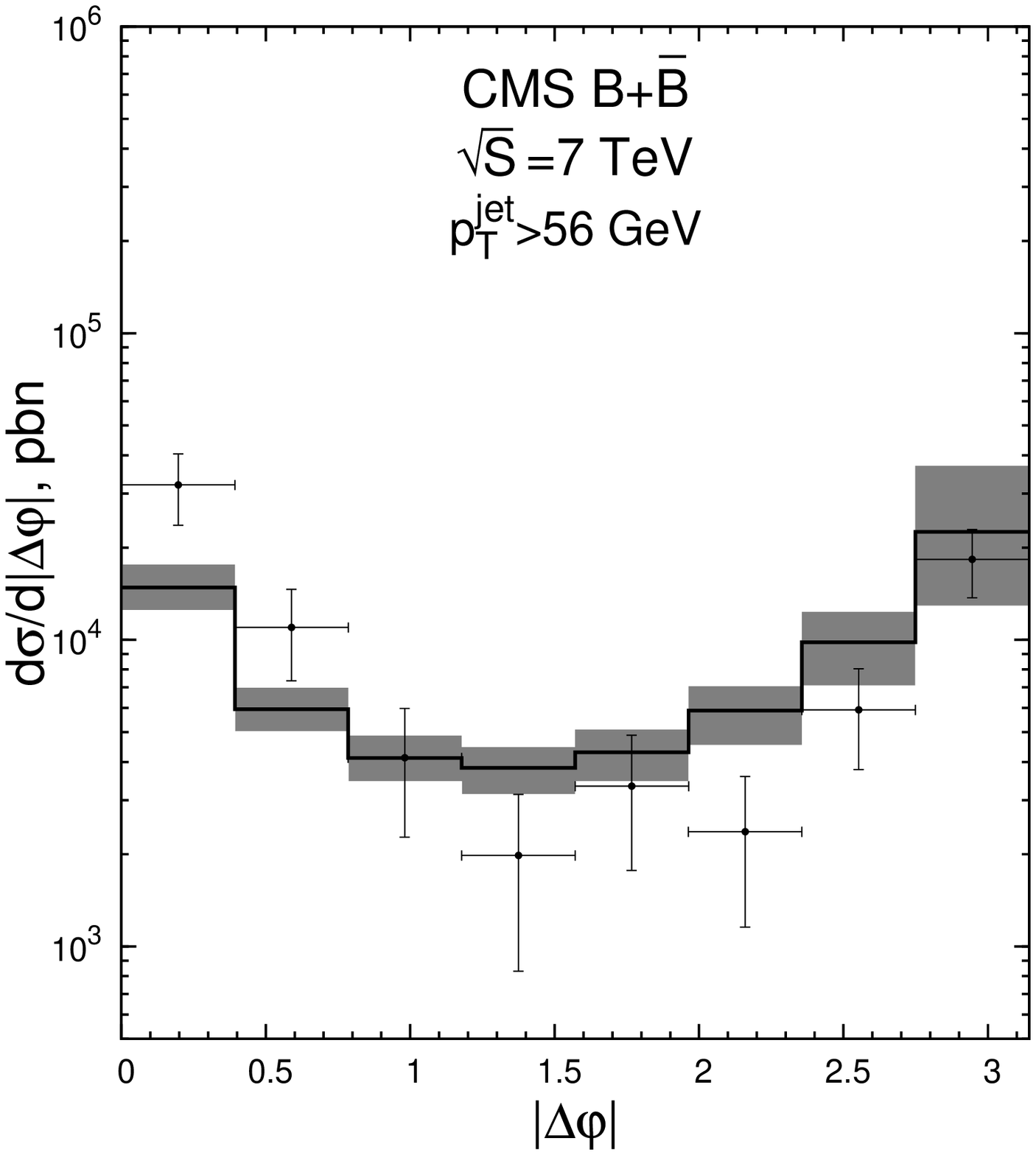}
\end{center}
\caption{Left panel: Diagrammatic representation of the
MRK-asymptotics of the squared amplitude of the subprocess
(\ref{eq:subp_fact}). The dashed lines denote Reggeized gluons
($R_\pm$). Part of the diagram, containing the interaction of the gluon
$g(p_2)$ with the Reggeized gluon $R_+(q_2)$ is not shown for brevity.
Right panel: the spectrum of $B\bar B$-meson pair as a funcion of
the azimuthal angle difference between them, the the data from CMS
Collaboration \cite{BBCMS}.
 \label{fig:factor}}
\end{figure}

  To derive the factorization formula of the LO PRA, let's consider the following auxiliary hard subprocess:
  \begin{equation}
  g(p_1)+g(p_2) \to g(k_1) + q\bar{q} (P_{\cal A}) + g(k_2), \label{eq:subp_fact}
  \end{equation}
  where the pair of heavy quarks $q=c/b$ with the four-momentum $P_{\cal A}$ and two additional gluons has been produced.
  In the Multi-Regge limit, when ${\bf k}_{T1}^2$, ${\bf k}_{T2}^2$ and ${\bf P}_{{\cal A}T}^2$ are $\ll \hat{s}=(p_1+p_2)^2$,
  and the rapidity gaps between produced particles $\Delta y(k_1,P_{\cal A}) \gg 1$ and $\Delta y(P_{\cal A}, k_2) \gg 1$,
  amplitude of the subprocess (\ref{eq:subp_fact}) is known to exhibit the $t$-channel factorization, presented diagrammatically
   in the Fig.~\ref{fig:factor} (left panel). This factorized form of the MRK amplitude is usually represented in terms of effective,
   gauge-invariant degrees of freedom of high energy QCD, Reggeized gluons ($R_{\pm}$) and Reggeized quarks ($Q_{\pm}$), which can be
   collectively denoted as {\it Reggeized partons}. Reggeized partons propagate in the $t$-channel between the effective vertices
   of the production of particles or clusters of particles, which are highly separated in rapidity. MRK amplitude of the process
   (\ref{eq:subp_fact}) has the following form:
  \begin{eqnarray}
  {\cal M} &=& g_s^2 \left(\Gamma_{\mu\nu -}(p_1,-k_1) f_{a_1 b_1 c_1} \varepsilon^{a_1}_\mu(p_1) \varepsilon^{b_1\star}_\mu(k_1) \right)
  \times \frac{1}{2 q^2_1}
 {\cal A}^{c_1c_2} \frac{1}{2 q^2_2} \nonumber \\
   &\times & \left(\Gamma_{\mu\nu +}(p_2,-k_2) f_{a_2 b_2 c_2} \varepsilon^{a_2}_\mu(p_2) \varepsilon^{b_2\star}_\mu(k_2) \right) \label{eq:MRK_ampl}
   \end{eqnarray}
where $g_s^2=4\pi\alpha_s$ is the squared coupling constant of QCD, $q_{1,2}=p_{1,2}-k_{1,2}$, ${\cal A}^{c_1 c_2}$ is the effective
vertex of the process $R_+R_-\to q\bar{q}$, and the effective $R_{\mp}gg$-vertex reads~\cite{LipatovRev, EffAct, KTAntonov}:
  \begin{equation}
  \Gamma_{\mu\nu\mp}(k_1,k_2)=2 g_{\mu\nu} k_1^\pm + (2k_2+k_1)_\mu n_\nu^\pm - (2k_1+k_2)_\nu n_{\mu}^\pm -
  \frac{(k_1+k_2)^2}{k_1^\pm}n_\mu^\pm n_\nu^\pm, \label{eq:Rgg-vert}
  \end{equation}
  where $k_{1,2}$ are (incoming) four-momenta of gluons. The vectors $n_+^\mu=P_2^\mu/\sqrt{S}$ and $n_-^\mu=P_1^\mu/\sqrt{S}$ are related
  with the four-momenta of colliding protons $P_{1,2}^\mu$, $P_{1,2}^2=0$, $S=2P_1P_2$. These vectors allows one to define the
  Sudakov (or light-cone) decomposition of the arbitrary four-vector:
  $k^\mu=\frac{1}{2}\left( n_+^\mu k^- + n_-^\mu k^+ \right) + k_T^\mu$,  where $n_\pm k_T=0$, $k_{\pm}=n_{\pm}k$ and $k_{\pm}=k^\pm$.

  Effective vertex (\ref{eq:Rgg-vert}) describes the interaction of a gluon, with large $k^+$ or $k^-$ light-cone momentum component with
  the Reggeized gluon $R_{\mp}$. The latter carries only one light-cone component of momentum $q^+$ for $R_+$ and $q^-$ for $R_-$, and therefore,
  the other light-cone components of momenta of the gluons is conserved: $k^\pm_1+k^\pm_2=0$ for the interaction with $R^\mp$ respectively.
  Taking this conditions into account, it is easy to see, that for the on-shell external gluons $k_{1,2}^2=0$ the vertex (\ref{eq:Rgg-vert})
  satisfies Slavnov-Taylor identities $k_1^\mu \varepsilon^\nu(k_2) \Gamma_{\mu\nu\mp}(k_1,k_2)=0$ and
  $\varepsilon^\mu(k_1) k_2^\nu \Gamma_{\mu\nu\mp}(k_1,k_2)=0$, which guarantees the gauge invariance of the $R_\pm gg$-scattering amplitude.

  The effective production vertex ${\cal A}^{c_1 c_2}$ is the Green's function of interaction of Reggeized gluons with ordinary QCD gluons and quarks,
   with amputated propagators of Reggeons. The effective vertex for the hard subprocess with the arbitrary number of quarks and gluons
   in the final state can be constructed, using the formalism of the Lipatov's effective action for Multi-Regge processes in QCD~\cite{EffAct, KTAntonov},
   and it will be {\it gauge invariant} for arbitrary values of virtualities of incoming Reggeons $q_{1,2}^2$, provided that the above-mentioned
   constraints on the light-cone components of momenta of the Reggeons $R_{\pm}$ and $Q_{\pm}$ are fulfilled.


Contracting the vertex (\ref{eq:Rgg-vert}) with polarization vectors of on-shell gluons, squaring it and summing over the helicities,
one obtains the following simple result:
  \begin{equation}
  \sum \limits_{\lambda, \lambda'} \left \vert \Gamma_{\mu\nu\pm}(k_1,-k_2) \varepsilon_\mu(k_1,\lambda)
  \varepsilon^\star_\mu(k_2,\lambda') \right \vert^2 = 8(k_1^\mp)^2, \label{eq:sqr-Rgg-vert}
  \end{equation}
  which allows one to rewrite the squared amplitude (\ref{eq:MRK_ampl}) summed (averaged) over the spin and color quantum numbers of the final-state (initial-state) particles in the following form:
  \begin{equation}
  \overline{|{\cal M}|^2}= \frac{4g_s^4}{q_1^2 q_2^2} \tilde{P}_{gg}(z_1) \tilde{P}_{gg}(z_2)
  \frac{\overline{|{\cal A}_{PRA}|^2}}{z_1z_2}, \label{eq:M2-final}
  \end{equation}
  where the squared PRA amplitude is defined as:
  \begin{equation}
  \overline{|{\cal A}_{PRA}|^2} = \left( \frac{q_1^+ q_2^-}{4(N_c^2-1) \sqrt{t_1t_2}} \right)^2
  \left[ {\cal A}^\star_{c_1c_2} {\cal A}^{c_1c_2}   \right]. \label{eq:PRA-A-pr}
  \end{equation}
  In the last expression we denote $t_{1,2}={\bf q}_{T1,2}^2$, also we have introduced the light-cone momentum fractions
  $z_1=q_1^+/p_1^+$, $z_2=q_2^-/p_2^-$ and the PRA gluon-gluon splitting function $\tilde{P}_{gg}(z)=2C_A(1-z)/z$.
  To derive Eqns. (\ref{eq:M2-final}) and (\ref{eq:PRA-A-pr}), one has to take into account, that $q_{1,2}^2=-{\bf q}_{T1,2}^2/(1-z_{1,2})$.

   Kinematically, the MRK limit $|\Delta y(k_{1,2},P_{\cal A})| \gg 1$ corresponds to the situation when $z_{1,2}\ll 1$,
   while the transverse momenta ${t}_{1}\sim {t}_{2}\sim |{\bf p}_{T{\cal A}}|$ can be non-negligible, and the effective
   production amplitude (\ref{eq:PRA-A-pr}) explicitly and non-trivially depends on $t_{1,2}$.

  The opposite limit $t_{1,2}\to 0$ in the Eq. (\ref{eq:M2-final}), corresponds to the traditional collinear factorization of QCD amplitudes.
  The collinear factorization for the squared QCD amplitude of the subprocess (\ref{eq:subp_fact}) is correctly reproduced by the
  Eq. (\ref{eq:M2-final}) in the region $z_{1,2}\ll 1$, since the following relation holds for the squared PRA amplitude (\ref{eq:PRA-A-pr}):
  \begin{equation}
  \int\limits_0^{2\pi}\frac{d\phi_1 d\phi_2}{(2\pi)^2} \lim\limits_{t_{1,2}\to 0} \overline{|{\cal A}_{PRA}|^2} =  \overline{|{\cal M}_{CPM}|^2},
  \end{equation}
  where $\overline{|{\cal M}_{CPM}|^2}$ is the squared amplitude of the corresponding subprocess with on-shell partons in the initial state
  (e. g. $gg\to q\bar{q}$ in the considered case), and PRA splitting function $\tilde{P}_{gg}(z)$ correctly reproduces the small-$z$ asymptotic
   of the full DGLAP splitting function $P_{gg}(z)=2C_A\left((1-z)/z + z/(1-z) + z(1-z)\right)$. One can observe, that to reproduce the collinear
    limit for any $z_{1,2}$, it is enough just to substitute the PRA splitting function in the Eq.~(\ref{eq:M2-final}) by the exact expression.

  Finally, we define the {\it modified MRK (mMRK) approximation} for the squared matrix element of the subprocess
  (\ref{eq:subp_fact}) by the Eq.~(\ref{eq:M2-final}) with the substitution $\tilde{P}_{gg}(z)\to P_{gg}(z)$,
  and we apply this approximation to all values of $t_{1,2}$ and $z_{1,2}$.  The numerical evidence (see Ref.~\cite{HEJ} for the case of Reggeized
  gluons and the Refs.~\cite{Hauptman, NS-2015} for the case of Reggeized quarks) suggests, that such approximation for matrix
   element is very good for the simulation of additional hard radiation outside of the final-state collinear region.

  To derive the factorization formula for the cross-section, we substitute the mMRK approximation for the matrix element
  (\ref{eq:M2-final}) to the standard formula of collinear factorization, integrated over the phase-space of additional partons:
  \[
  d\sigma = \int \frac{dk_1^+ d^2{\bf k}_{T1}}{2k_1^+} \int \frac{dk_2^- d^2{\bf k}_{T2}}{2k_2^-} \int\limits_0^1
  d\tilde{x}_1 d\tilde{x}_2\ f_{g}(\tilde{x}_1,\mu^2) f_{g}(\tilde{x}_2,\mu^2)\cdot \frac{\overline{|{\cal M}|^2}}{2S\tilde{x}_1
  \tilde{x}_2} d\Phi_{\cal A},
  \]
  where $f_g(x,\mu^2)$ are the standard Parton Distribution Functions of CPM, $\tilde{x}_{1}=p^+_{1}/P^+_{1}$,
  $\tilde{x}_{2}=p^-_{1}/P^-_{1}$, and $d\Phi_{\cal A}$ is the element of Lorentz-invariant phase space for the partons
  in the final state of the hard subprocess. Finally, one can change the variables of integration in the last expression as follows:
  $(\tilde{x}_1, k_1^+, {\bf k}_{T1})\to (x_1, z_1, {\bf q}_{T1})$ and $(\tilde{x}_2, k_2^-, {\bf k}_{T2})\to (x_2, z_2, {\bf q}_{T2})$,
  where the variables $x_{1,2}$ are defined as $x_1=q_1^+/P_1^+$, $x_2=q_2^-/P_2^-$, and rewrite it in a following form:
  \begin{equation}
  d\sigma = \int\limits_0^1 \frac{dx_1}{x_1} \int \frac{d^2{\bf q}_{T1}}{\pi} \tilde{\Phi}_g(x_1,t_1,\mu^2) \int\limits_0^1
  \frac{dx_2}{x_2} \int \frac{d^2{\bf q}_{T2}}{\pi} \tilde{\Phi}_g(x_2,t_2,\mu^2)\cdot \frac{\overline{|{\cal A}_{PRA}|^2}}{2Sx_1x_2}
   d\Phi_{\cal A}, \label{eq:kT-fact}
  \end{equation}
  where the ``tree-level'' unintegrated PDFs (unPDFs) have the form:
  \begin{equation}
  \tilde{\Phi}_{g}(x,t,\mu^2)=\frac{\alpha_s}{(2\pi)} \int\limits_x^{1} \frac{dz}{t}\ P_{gg}(z)\cdot \frac{x}{z} f_g
  \left( \frac{x}{z},\mu^2 \right).\label{eq:tree-Phi}
  \end{equation}

  The Eq.~(\ref{eq:kT-fact}) is nothing but the well-known formula of the $k_T$-factorization of the cross-section~\cite{kT_fact},
  but integrals over $t_{1,2}$ and $z$ in the Eqns.(\ref{eq:kT-fact}) and (\ref{eq:tree-Phi}) are logarithmically divergent for $t\to 0$ and $z\to 1$.
   This divergence can be regulated if one properly takes into account the leading doubly-logarithmic corrections
   $\sim \left( \alpha_s \log^2 (t/\mu^2)\right)^n$ from all orders of perturbation theory~\cite{DDT}.
   Here we follow the approach of Kimber, Martin and Ryskin (KMR)~\cite{KMR}, where the unPDF is constructed to satisfy the normalization condition:
  $  \int\limits_0^{\mu^2} dt\ \Phi_i(x,t,\mu^2) = x f_i(x,\mu^2)$,
  which ensures approximate normalization of the results for single-scale observables, obtained in $k_T$-factorization,
   on the corresponding LO CPM results. In the KMR approach, the $z\to 1$-singularity is regulated by the condition
    of rapidity ordering of the last emitted parton with the particles, produced in the hard subprocess.
    The last condition is natural from the point of view of our mMRK approximation, since only in this kinematic region it provides
     good approximation for the exact QCD matrix element.

  The last important conceptual point is related with the Eq. (\ref{eq:kT-fact}). As a result of the derivation, presented above,
  the flux factor for the off-shell initial-state partons with virtualities $t_{1,2}$ is shown to coincide with the flux factor,
  which one has in CPM: $2Sx_1x_2=(2S\tilde{x}_1\tilde{x_2})\cdot z_1z_2$. This result holds both in the collinear and in the Regge limits.
  Also this flux factor is shown to be compatible with the KMR unPDFs, since the Eq.~(\ref{eq:tree-Phi}) reproduces the ``tree-level''
  structure of the KMR unPDFs (without Sudakov formfactor), and the flux factor is purely kinematical quantity,
  which could be determined on a basis of consideration of tree-level amplitudes only, without any need to take into account the loop corrections.

\section{Heavy quark production at the LHC\label{sec:heavy}}

\subsection{Production of $b\bar b-$jet pairs in the LO
PRA\label{sec:bjet}} 
\begin{figure}[h]
\begin{center}
\includegraphics[width=.4\textwidth, clip=]{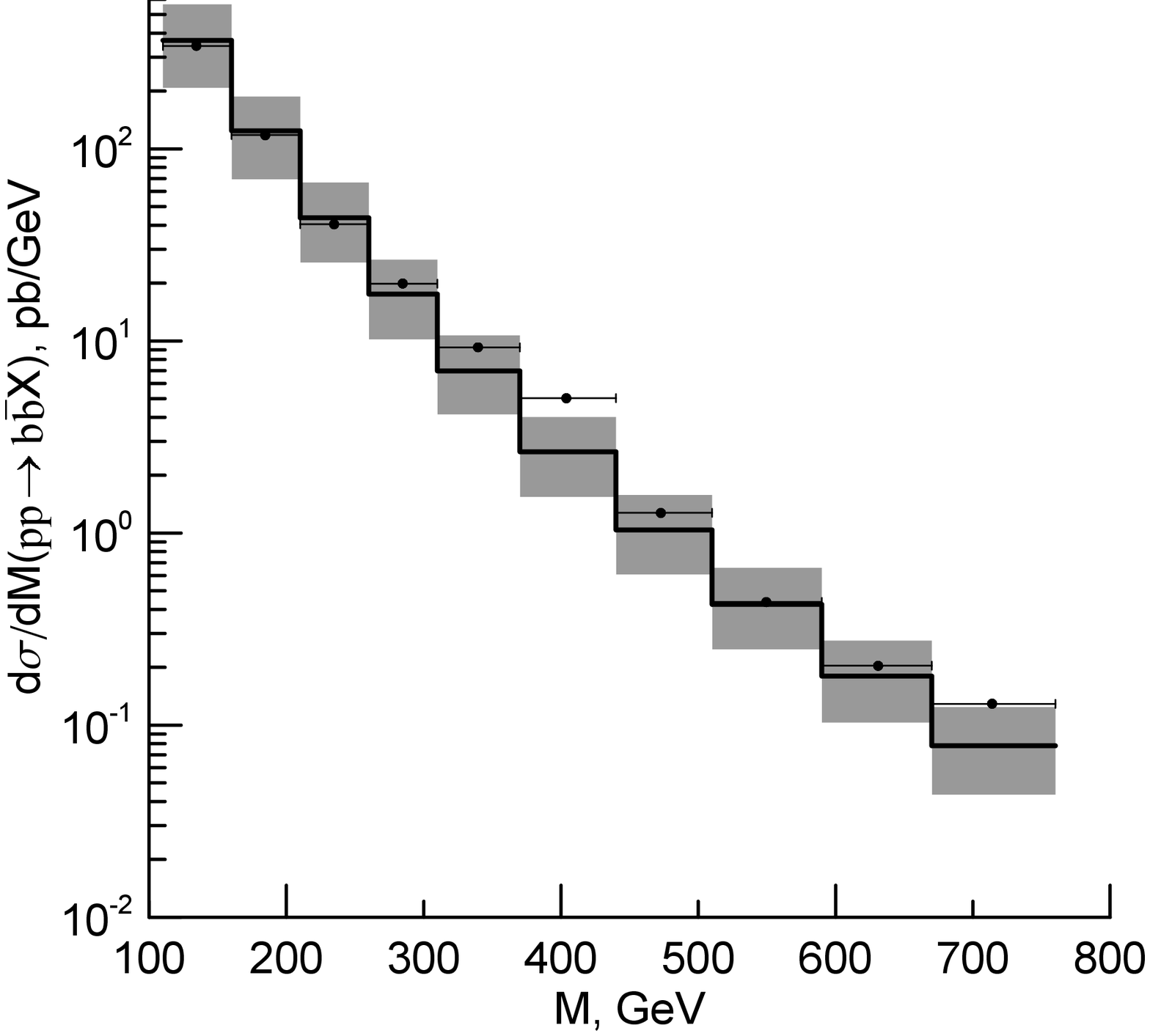}\includegraphics[width=.4\textwidth, clip=]{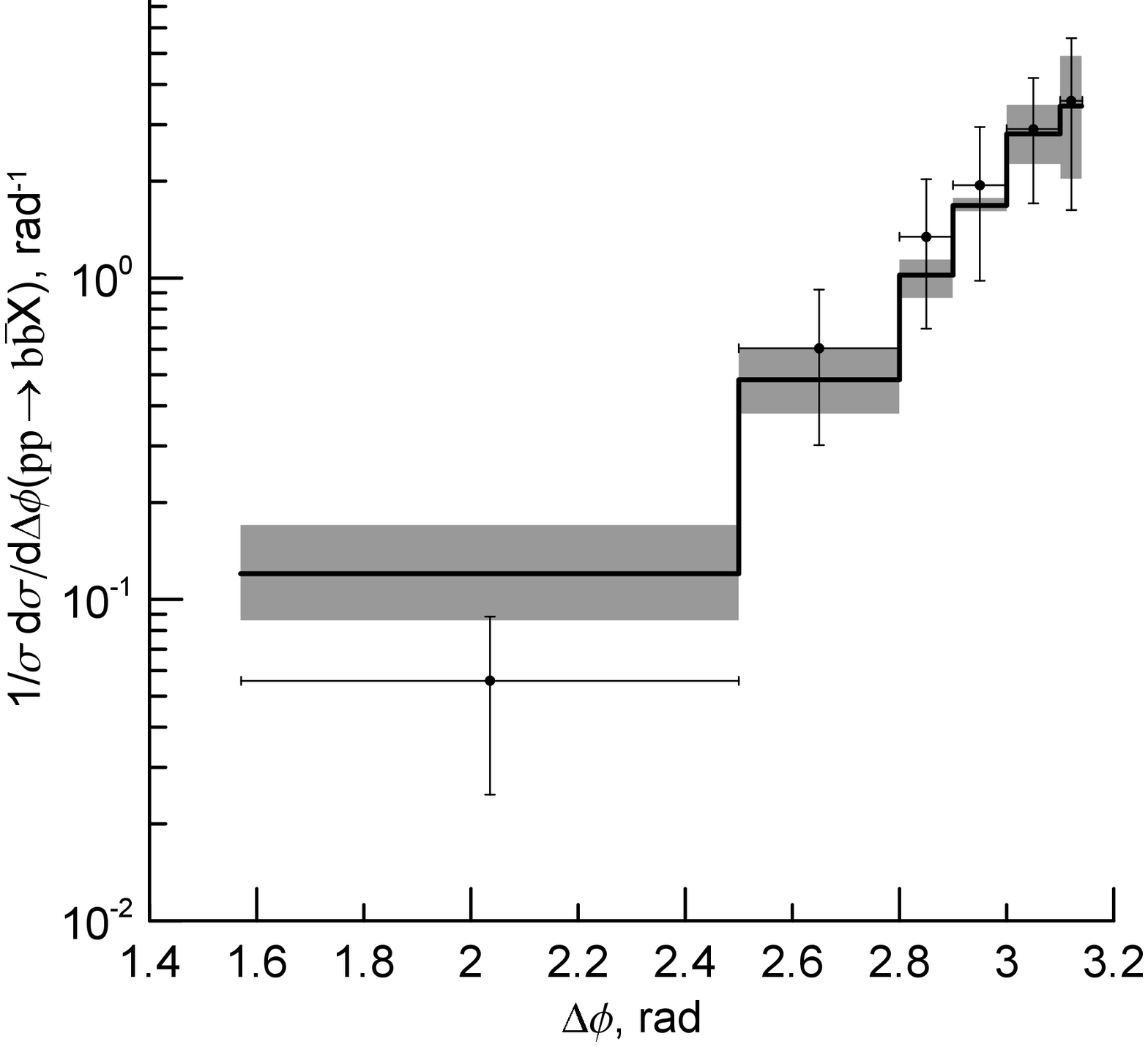}
\end{center}
\caption{\label{fig:2} The $b\bar b$-dijet cross-section as a
function of dijet invariant mass $M$ and as a function of the
azimuthal angle difference between the two jets for $b$-jets with
$p_T>40$~GeV, $|y|<2.1$. The data are from ATLAS Collaboration
\cite{ATLASb}, the solid line corresponds to KMR unPDF, the
shaded bands indicate the theoretical uncertainties.}
\end{figure}

\begin{figure}[h]
\begin{center}
\includegraphics[width=.4\textwidth, clip=]{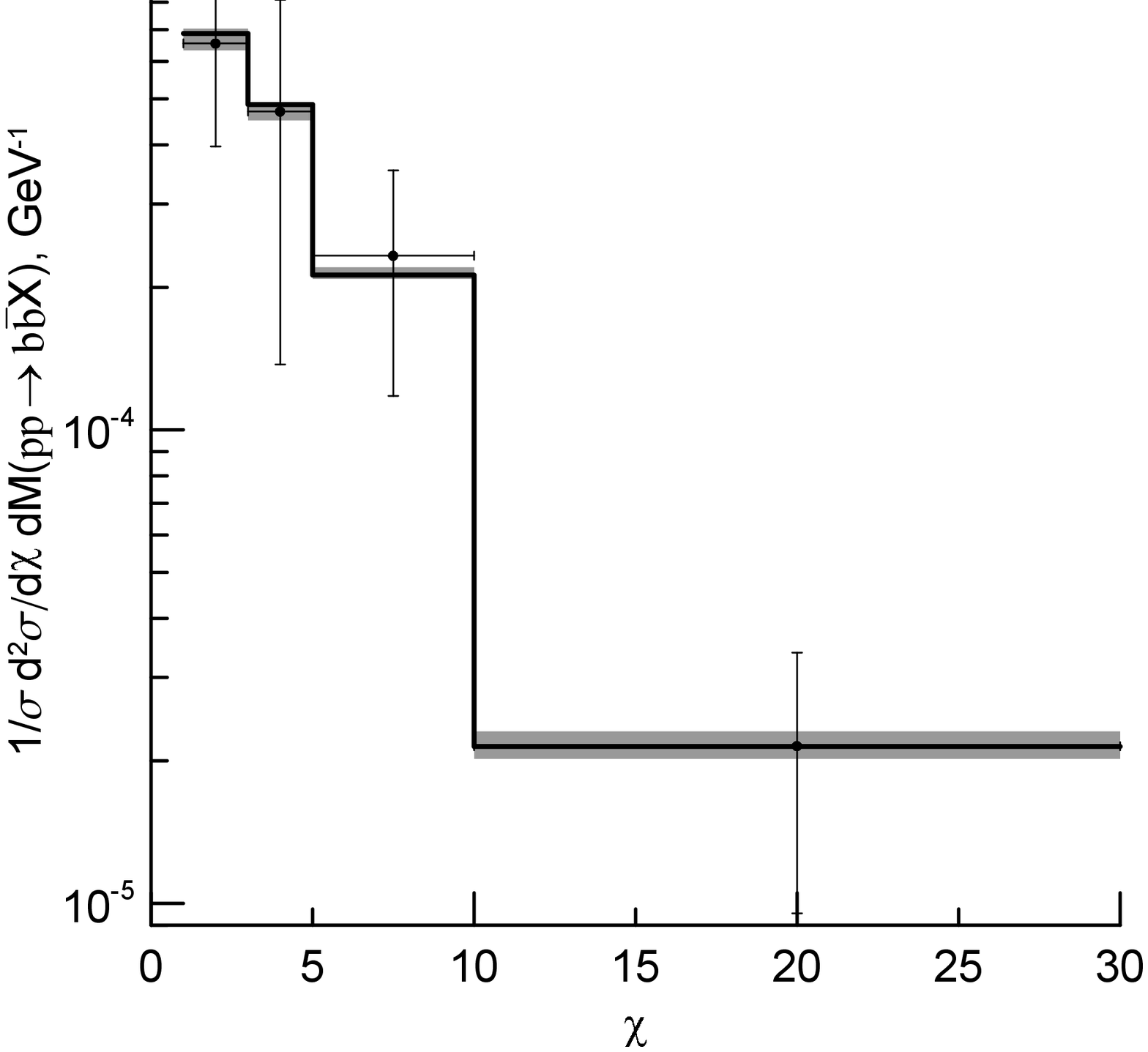}\includegraphics[width=.4\textwidth, clip=]{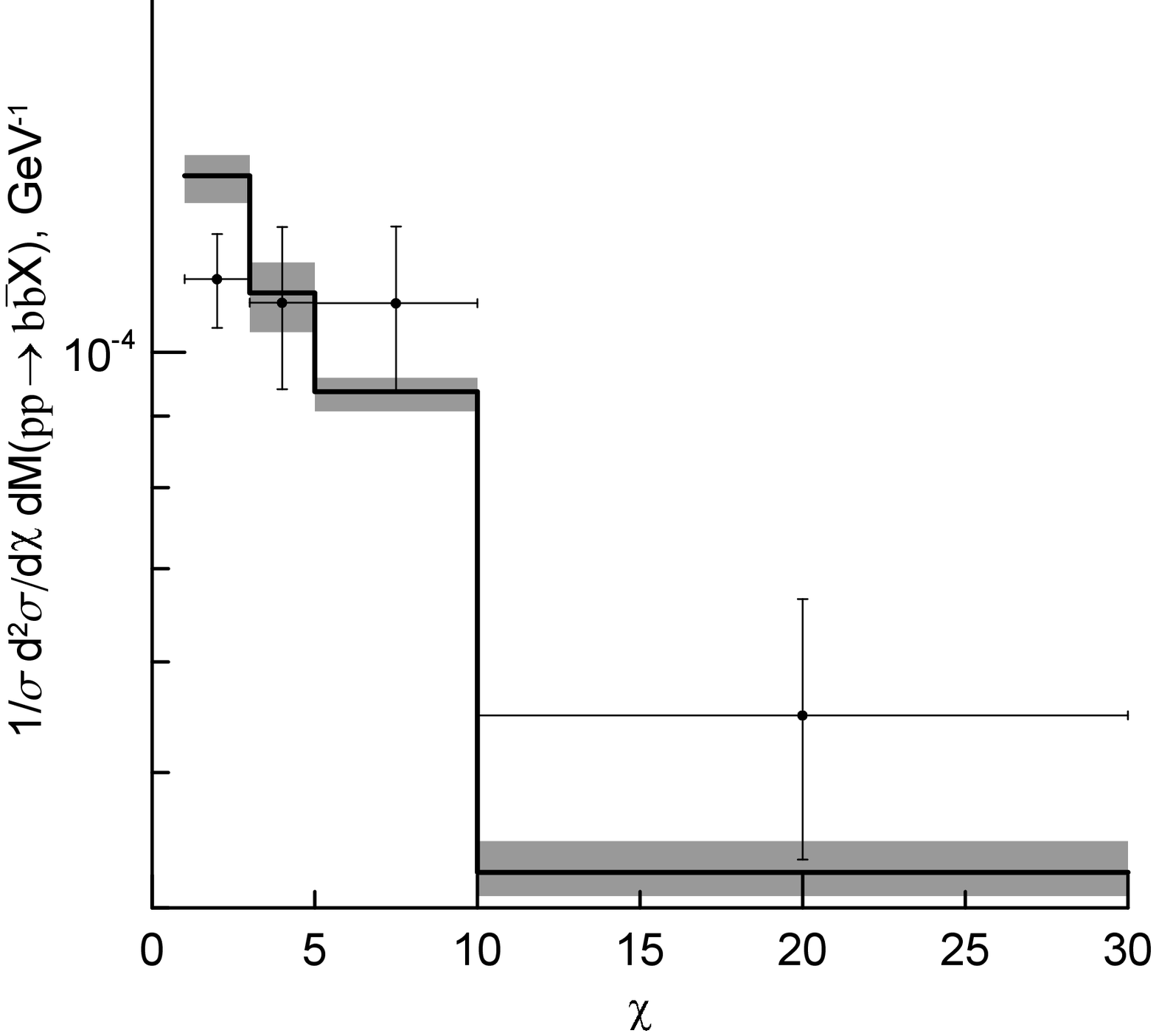}
\end{center}
\caption{\label{fig:3} The $b\bar b$-dijet cross-section as a
function of $\chi$ for $b$-jets with $p_T>40$~GeV, $|y|<2.1$ and
$|y_{boost}|=\frac{1}2|y_1+y_2|<1.1$, for dijet invariant mass range
$110<M<370$~GeV  and $370<M<850$~GeV. The data are from ATLAS
Collaboration \cite{ATLASb}.}
\end{figure}

To describe inclusive $b\bar b$-jet cross
section in the LO approximation of the PRA, we need to consider
gluon fusion subprocesses of  $b\bar b$-quark pair production, which
is to be dominant at the high energy, i.e. $ R_+ +R_- \to b+\bar b$. The
amplitude of this process is obtained accordingly Feynman rules of
Lipatov's effective theory \cite{KTAntonov} and the squared amplitude
can be taken in analytical form from Ref.~ \cite{NSS2013}.
In the Figs. \ref{fig:2} and
\ref{fig:3}, we demonstrate good agreement between LO PRA and data
from ATLAS Collaboration \cite{ATLASb} for invariant mass, azimuthal
angle difference and $\chi=\exp |y_1-y_2|$ spectra of $b\bar
b-$jets. The calculation was done using KMR model for unPDF
\cite{KMR}, which where obtained from the LO MSTW-2008 PDF set~\cite{MSTW2008}.

\subsection{Production of $D\bar D(DD)-$meson pairs in the LO
PRA\label{sec:dmeson}}

To describe the hadronization stage we should use the fragmentation
model, in which transition from the produced in hard interaction
parton $i$ to the $D(\bar D)$ meson is described by corresponding
fragmentation functions $F_{i\to D(\bar D)}(z,\mu^2)$ at the scale
$\mu^2$. In the case of $D\bar D(DD)$-meson pair production the
fragmentation formula has the following form:
\begin{eqnarray}
&&\frac{d\sigma(p+p\to D+\overline{D} + X)}{dp_{DT} dy_D
dp_{\overline{D}T} dy_{\overline{D}}}= \sum_{ij} \int_0^1
\frac{dz_1}{z_1} \int_0^1\frac{dz_2}{z_2} F_{i\to D}(z_1,\mu^2) F_{j\to \overline{D}}(z_2,\mu^2)\times \nonumber \\
&&\times\frac{d\sigma(p+p\to i(k_i=p_D/z_1)+j(k_j=p_{\overline{D}
(D)}/z_2)+ X)}{dk_{iT}dy_i dk_{jT}dy_j},\label{eq:fragDD}
\end{eqnarray}
where subprocesses $R_+ +R_- \to c+\bar c$ and $R_+ +R_- \to g+g$ contribute in
$D\bar D$-production, and subprocess $R_+ +R_- \to g+g$ contributes in
$DD$-production. In our calculations we use the LO FFs from
Ref.~\cite{KKSS}. These FFs satisfy two desirable properties: at
first, their scaling violations are ruled by DGLAP evolution
equations; at second, they are universal.

\begin{figure}[h]
\begin{center}
\includegraphics[width=0.4\textwidth, angle=-90,origin=c, clip=]{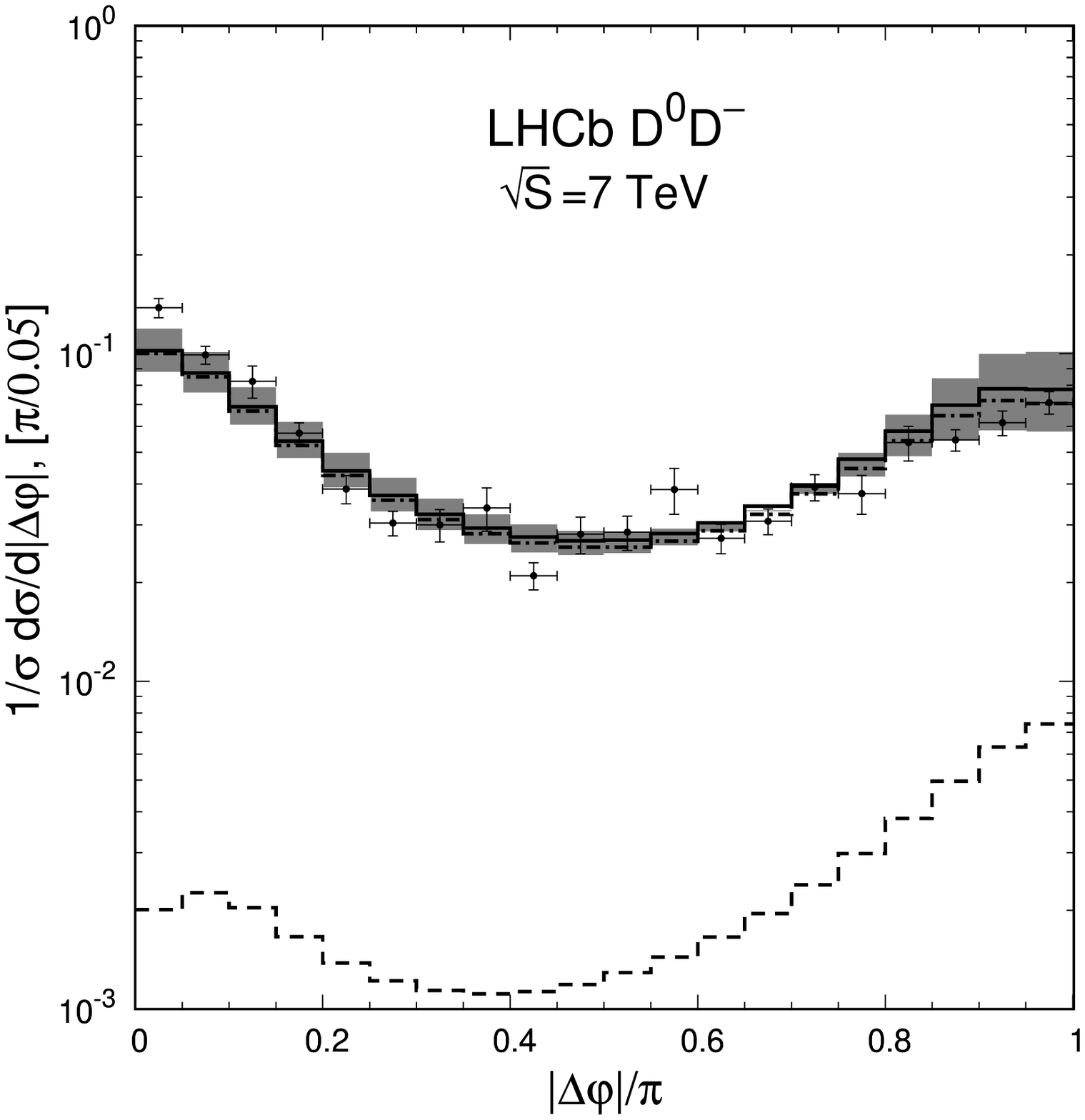}\includegraphics[width=0.4\textwidth, angle=-90,origin=c, clip=]{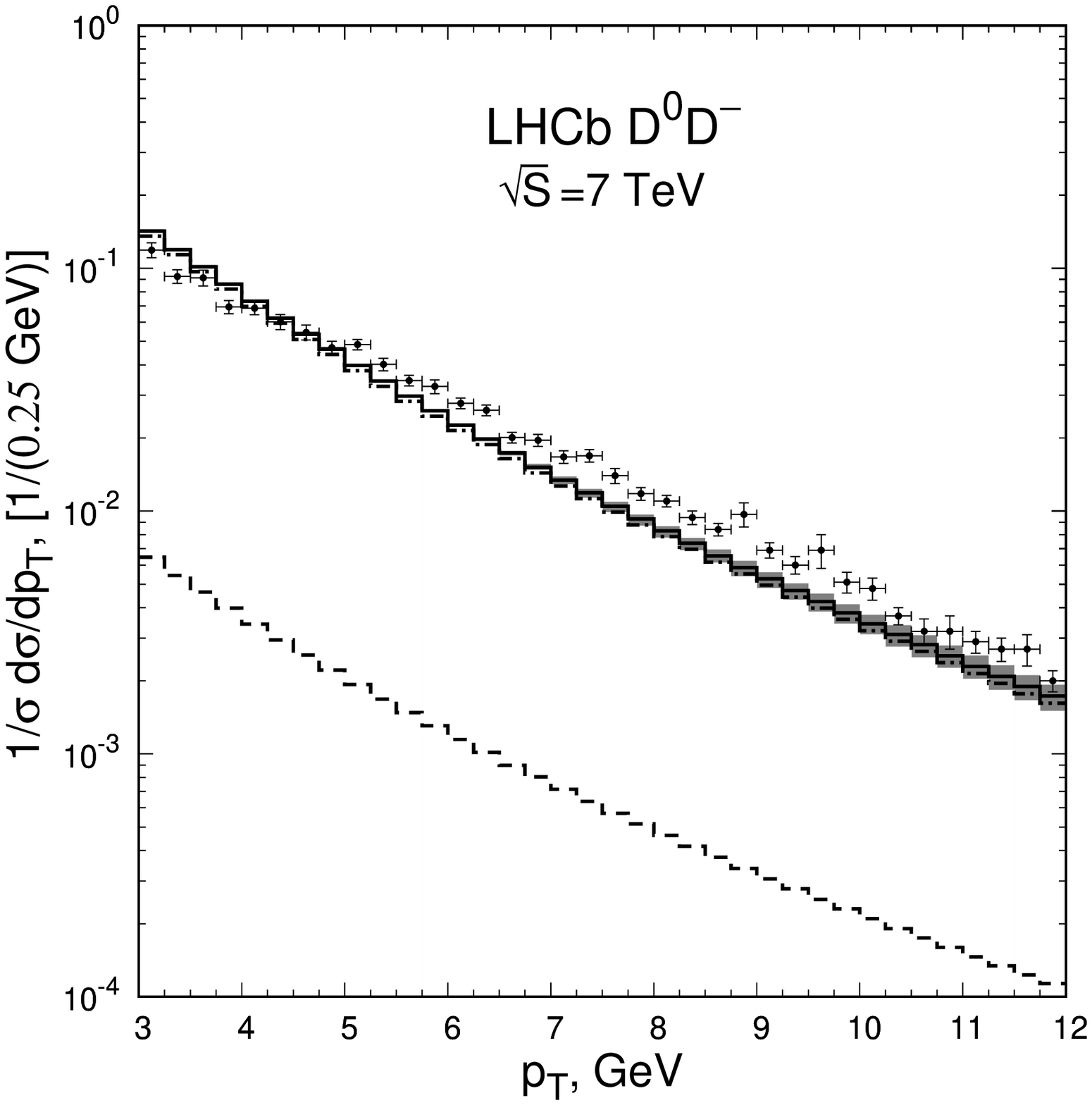}
\includegraphics[width=0.4\textwidth, angle=-90,origin=c, clip=]{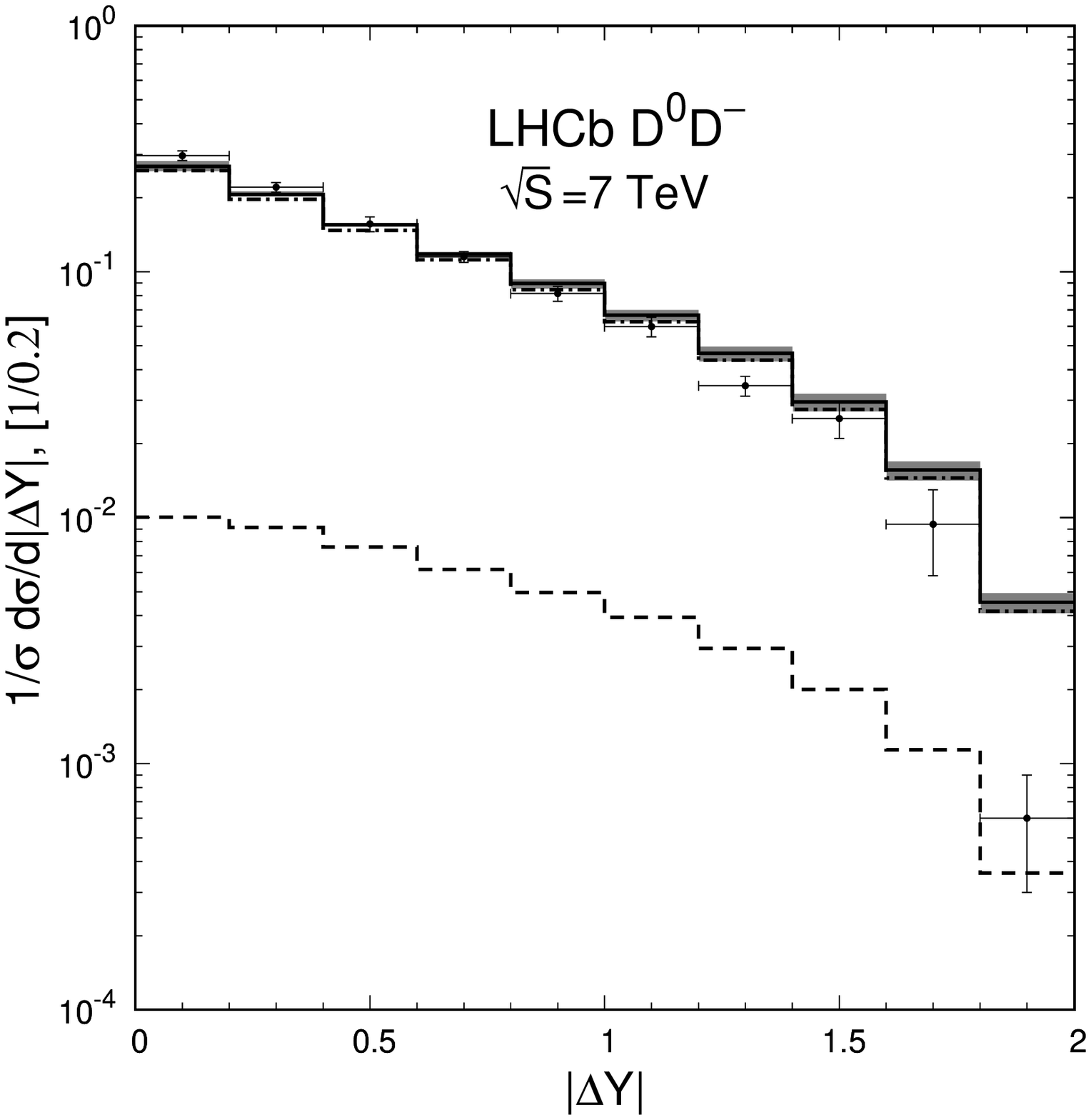}\includegraphics[width=0.4\textwidth, angle=-90,origin=c, clip=]{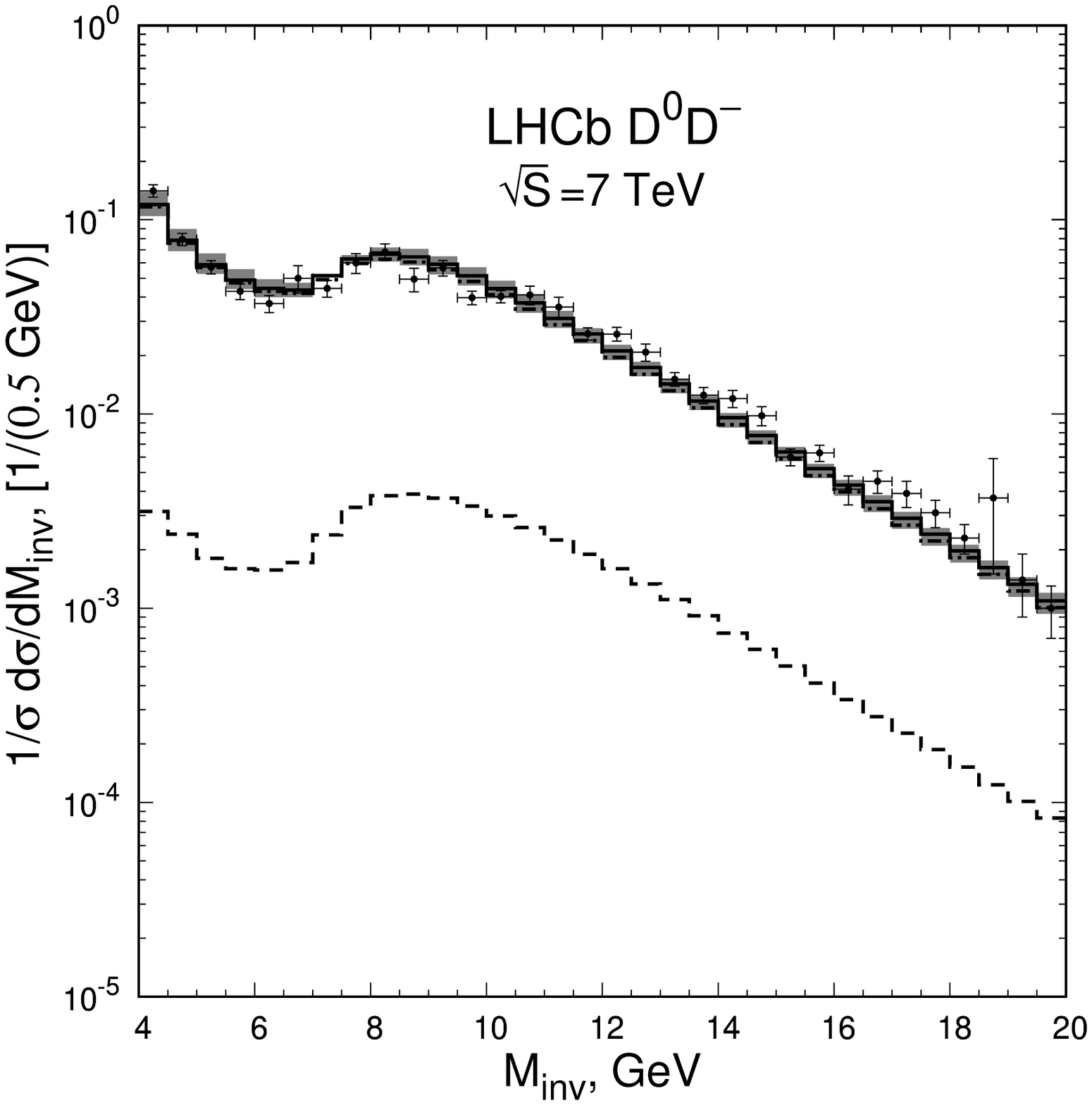}
\caption{The spectra of $D^0D^-$ pairs differential in azimuthal
angle difference (left, top), transverse momentum (right, top),
rapidity distance (left, bottom) and invariant mass of the pair
(right, bottom) at the $2<y<4$ and $\sqrt{S}=7$~TeV. The LHCb data
at LHC are from the Ref.~\cite{LHCb_Pair}. Dashed line represents
the contribution of gluon fragmentation in gluon-gluon fusion,
dash-dotted line -- the $c$-quark fragmentation contribution in
gluon-gluon fusion, solid line is their sum.\label{fig:4}}
\end{center}
\end{figure}

\begin{figure}[h]
\begin{center}
\includegraphics[width=0.4\textwidth, angle=-90,origin=c, clip=]{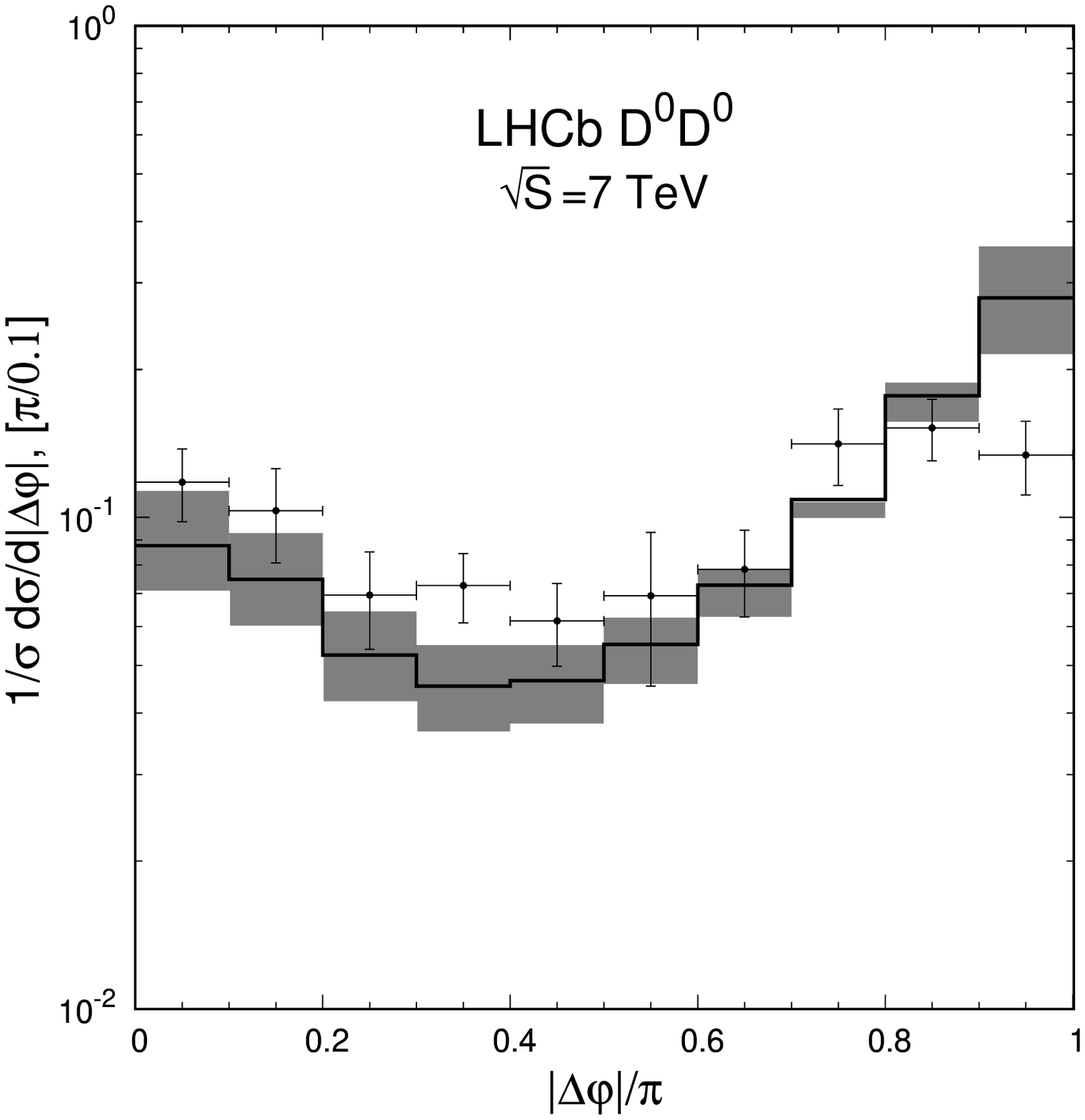}\includegraphics[width=0.4\textwidth, angle=-90,origin=c, clip=]{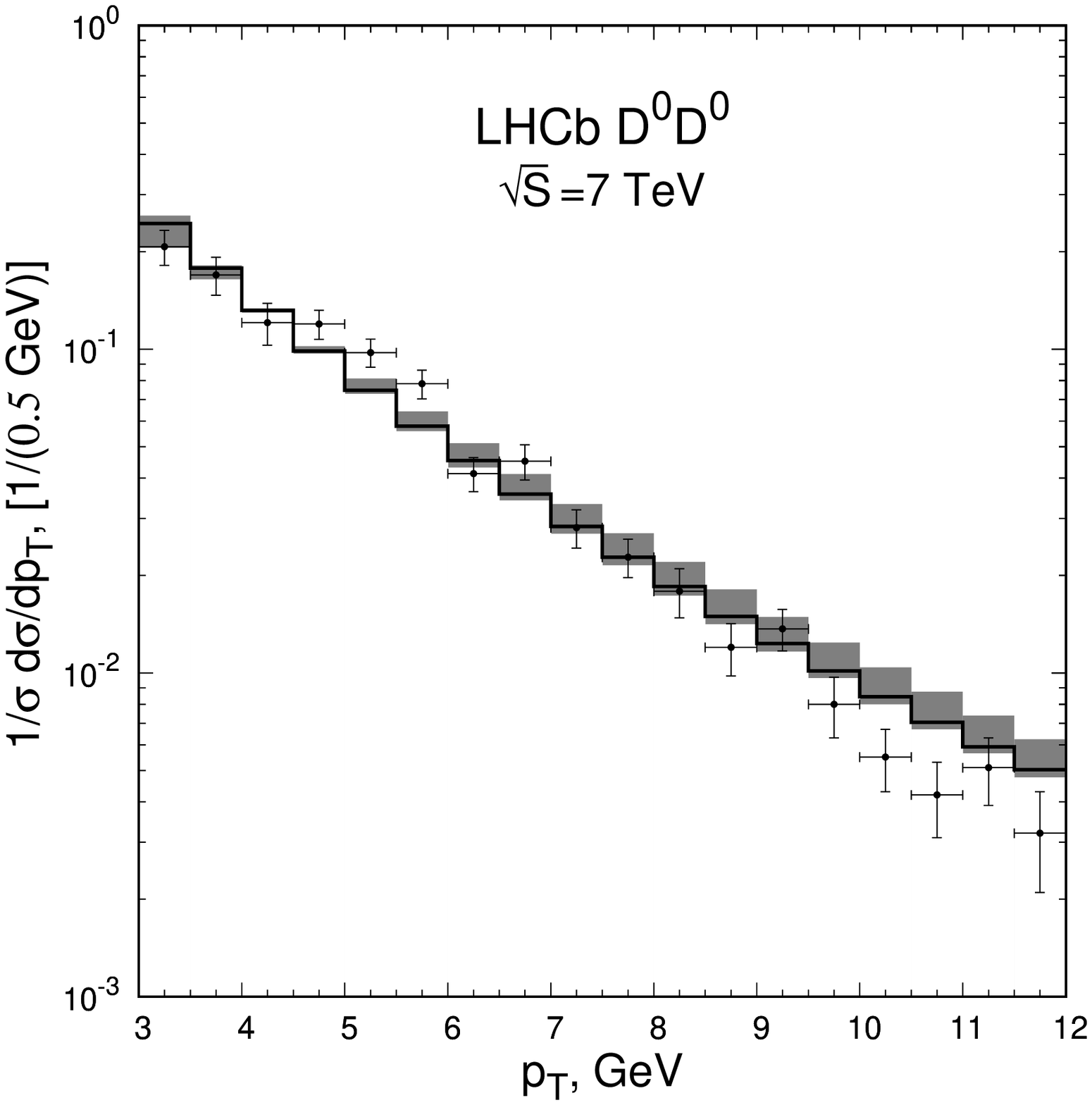}
\includegraphics[width=0.4\textwidth, angle=-90,origin=c, clip=]{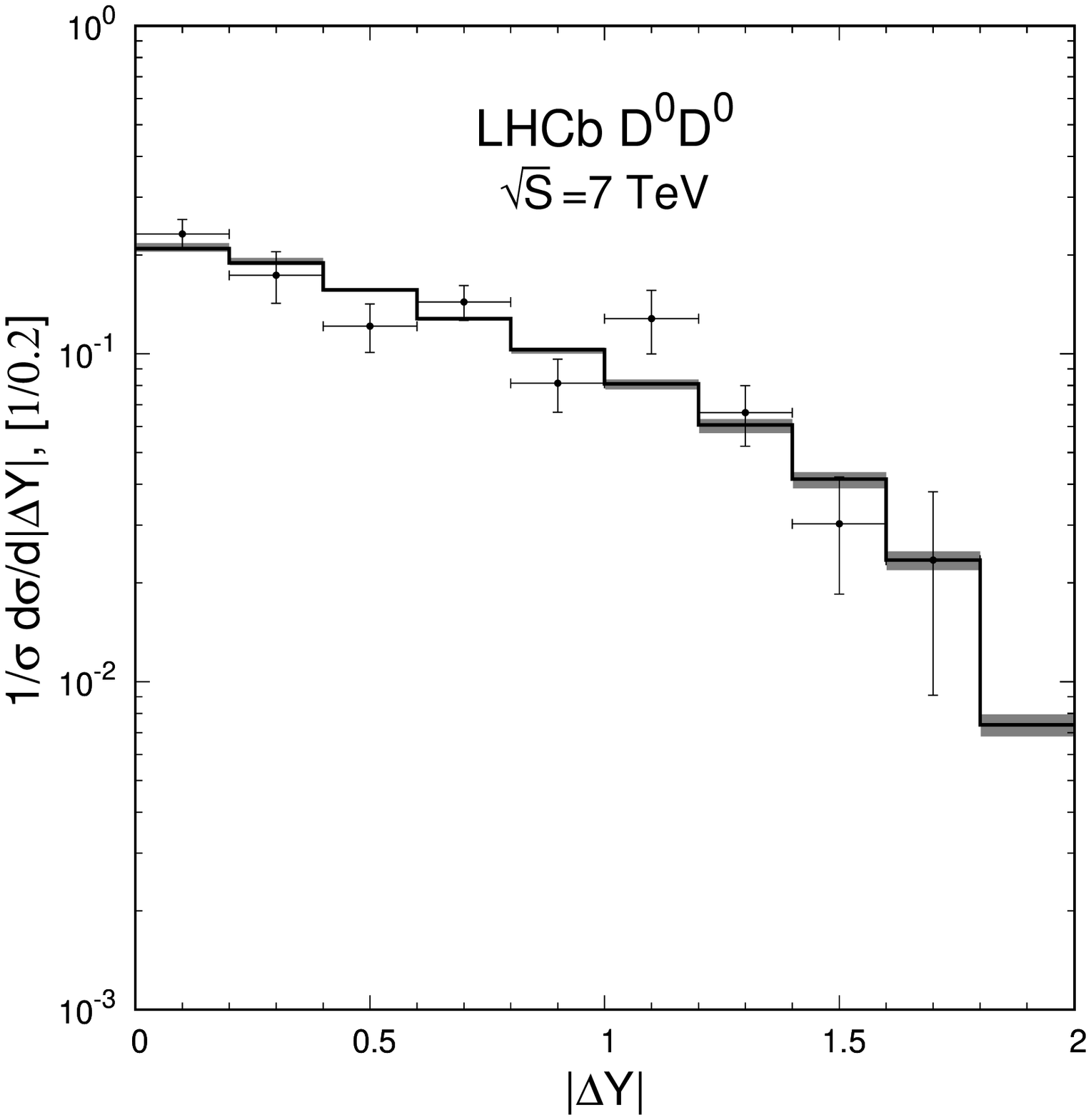}\includegraphics[width=0.4\textwidth, angle=-90,origin=c, clip=]{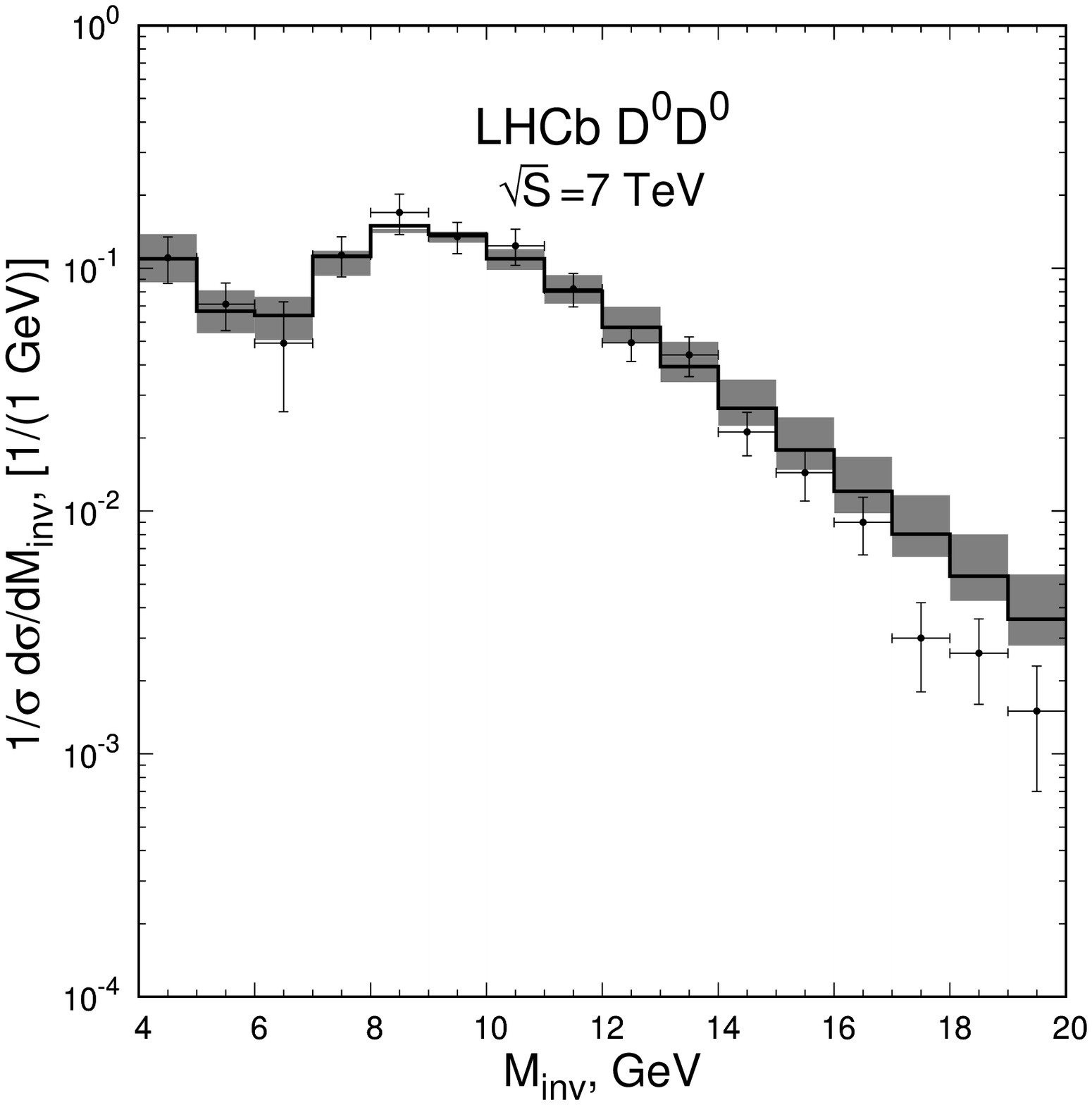}
\caption{The spectra of $D^0D^0$ pairs differential in azimuthal
angle difference (left, top), transverse momentum (right, top),
rapidity distance (left, bottom) and invariant mass of the pair
(right, bottom) at the $2<y<4$ and $\sqrt{S}=7$~TeV. The LHCb data
at LHC are from the Ref.~\cite{LHCb_Pair}. Solid line represents the
leading contribution of gluon fragmentation in gluon-gluon
fusion.\label{fig:5}}
\end{center}
\end{figure}

In Fig. \ref{fig:4}, the predicted in the LO PRA spectra of $D^0D^-$
pairs differential in azimuthal angle difference, transverse
momentum, rapidity distance and invariant mass of the pair are shown
in comparison with LHCb Collaboration data \cite{LHCb_Pair}. We see
that description of these two-particle spectra match better than it
is obtained in the NLO calculations of CPM in all kinematical
regions (see the comparison between theory and data in
\cite{LHCb_Pair}). The similar result is obtained in case of
$B\bar B-$meson pair production at the LHC, as it is shown in the
right panel of Fig. \ref{fig:factor}. In case of production of two
$D$ mesons, we also obtain good agreement with the data (see Fig.
\ref{fig:5}), taking into account gluon to $D-$meson fragmentation
mechanism. Let us note that $DD-$meson pair production can not be
described at all in the LO single-parton-scattering approximation of CPM
and the double-parton-scattering (DPS) production mechanism has been
suggested to describe these data \cite{DPS}. However, the
kinematical region where DPS dominates coincides with the domain
where the PRA gives large additional contribution to the CPM. As we
have obtained, there is no place for DPS contribution if we are
working already in the LO of PRA.

\section*{Acknowledgements}
The work was supported by Russian Foundation for Basic Research
through the Grant No~14-02-00021, and by the Ministry of Education
and Science of Russia under Competitiveness Enhancement Program of
Samara University for 2013-2020.


\begin{footnotesize}


%
\end{footnotesize}


\end{document}